\documentclass[aps,showpacs,preprintnumbers,amsmath,amssymb,twocolumn, tightenlines]{revtex4}

\usepackage{graphicx}
\usepackage{bm}
\usepackage{amsmath}
\usepackage{dcolumn}
\usepackage{bm}

\newcommand{\be}{\begin{eqnarray}}
\newcommand{\ee}{\end{eqnarray}}

\begin{document}
\title{Macroscopic Chirality Fluctuations in Heavy Ion Collisions \\ should induce CP forbidden Decays }

\author{ Raffaele Millo$^1$ and Edward V. Shuryak$^2$ }

\affiliation{ $^1$Universit\'a degli Studi di Trento and I.N.F.N.\\ Via Sommarive 14, Povo (Trento), Italy.\\
$^2$Department of Physics and Astronomy, State University of New York, 
Stony Brook NY 11794-3800, USA
}

\date{\today}

\vspace{0.1in}
\begin{abstract}
If large  fluctuations of quark chirality occur in heavy ion collisions,  they result in macroscopic CP-odd ``spots" of the so called theta-vacua,
with a non-zero $\theta(x)$. We consider particular decays of mesons, CP-forbidden in the vacuum with zero $\theta$, like $\eta\rightarrow \pi\pi$. We
 evaluate their rates for  such decays near hadronic freezeout. These rates, as well as charge asymmetries already observed, are
proportional to square of the CP-violating parameter $\langle\theta^2\rangle$ averaged over the fireball and events. 
With such input, we 
  found that the forbidden decay  rates are likely to be orders of magnitude larger than CP-allowed
ones. We further estimated that 
  up to about one per mill of $\eta$ mesons produced in heavy ion collisions
should decay in this way. We further discuss how those can be observed. We argue using STAR data on  charge asymmetries for AuAu and CuCu
collisions that the size of CP-odd spots at freezeout is as large as Cu nuclei: this fortunate fact (not explained so far by itself) suggests that the two-pion
invariant mass is modified by only about a percent, which is comparable to experimental resolution. If so, we think
experimental observation of these decays  is within the reach of current dataset. If those decays are found, it would confirm
that CP-odd interpretation of charge asyymetry is correct, even without complication related to geometry, impact parameter or magnetic field
induced on the fireball.
\end{abstract}

\maketitle

\section{Introduction}
  Discovery of instantons \cite{Belavin:1975fg} have revealed the dynamical role of gauge field topology, in particular existence of the $\theta$-vacua and
  the so called strong CP problem. Without going into discussion of its cosmological aspects and axions, let us only mention how it plays out in a ``Little Bang",
  the heavy ion collisions.
   One notable suggestion, by 
  Kharzeev, Pisarsky and Tytgat \cite{Kharzeev:1998kz}, was that analogs of theta-vacua can 
   possibly appear at finite temperature, as  metastable CP-odd bubbles. This might happened if the $\eta'$ mass can get very 
  small  near $T_c$. However, lattice measurements of the topological susceptibility near $T_c$, such as e.g.
\cite{Lucini:2004yh}, seem to exclude this scenario. But there is no need to have metastability of the bubbles: in fact initial state fluctuations of chirality
would evolve in diffusively, as it is preserved in the QGP. The strongly coupled nature of it, manifesting itself in small viscosity and diffusion
constant for charm, only helps to keep the CP-odd spot localized.

But before we get to topological effects in heavy ion collisions, let us discuss the QCD vacuum first.
The first empirical information on instantons in the QCD  vacuum has been provided long ago by one of us \cite{Shuryak:1981ff}: based on the values of gluon and quark condensates
it was conjectured that the mean size and the density  (instantons plus anti-instantons together) are
\be \rho_{inst}\approx 0.3 \, fm,\,\, n_{inst}\approx 1 \,\, fm^{-4}\ee
These numbers were confirmed by direct lattice observations a decade later. For details of that and for summary on QCD instanton-induced effects in chiral symmetry breaking
and hadronic spectroscopy  see \cite{Schafer:1996wv}.

The role of instantons in high energy collisions have been discussed in \cite{Kharzeev:2000ef},\cite{Shuryak:2000df}, in which it was suggested that instantons
are responsible for the so called ``soft pomeron", operating at a momentum scale $\sim 1/\rho\sim 1 \, GeV$, less than that for perturbative BFKL Pomeron. 
Eventually it lead to explanation of how
  topology/chirality fluctuations should occur in high energy collisions. It has been found 
that instantons, being perturbed by a collision, are excited and explode into certain states of glue and fermions which can be described semiclassically.
Transition from virtual (Euclidean) to real (Minkowskian)
part of the semiclassical path proceed via the so called ``turning states" , at which the momentum (electric field) vanishes . Such ``turning states"
have been identified as the so called Carter-Ostrovsky-Shuryak (COS)  
  sphalerons \cite{Ostrovsky:2002cg}.  Analitic solutions for them, as well as classical description of their subsequent decay have been made in
  the same paper. Further 
 clarification of the fermion production in sphaleron explosion has been made by Shuryak and Zahed in \cite{Shuryak:2002qz}, by solving Dirac equation in the exploding
sphaleron background. It clearly shows how chirally asymmetric set of quarks $\bar u_R u_L  \bar d_R d_L \bar s_R s_L $ are pulled from
the Dirac sea into the positive energy states,  violating chirality by 6 units (as anomaly/index theorem equation predicts), with the calculated spectrum. 
It has been proposed that one can identify and experimentally study COS sphaleron production in double-diffractive pp collisions, and currently there is
some proposal to perform such experiments at RHIC.

\begin{figure}[!ht]
\vspace{0.75cm}
\includegraphics[width=4cm]{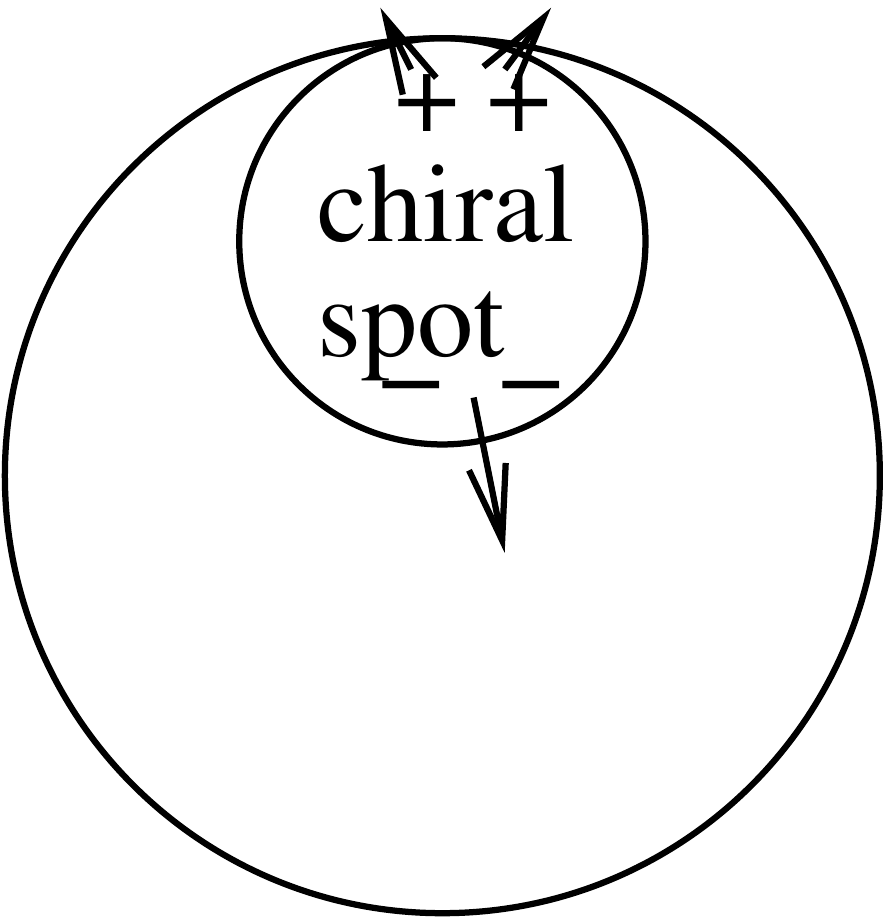}
\caption[h]{\label{fig_spot} A sketch of a chirally asymmetric spot (inner circle) in a fireball at freezeout. 
The external magnetic field (not shown) is assumed to be directed vertically: induced electric
field leads to charge separation (plusses upwards, minuses downwards). 
The spot's location is biased upward, as explained in the text. During the decays,  
the $++$ correlation is quenched by ambient matter less than
$+-$.}\label{fig1}
\end{figure}

Returning to Au-Au collisions, one may use these ideas in order to
estimate  the number of sphalerons produced  \cite{Shuryak:2003sb}. If we only consider
the number of vacuum instantons in an initial state ``pancake" then we get a so  minimal number
\be N_{sphalerons} >  n_{inst} \rho_{inst}^2 \pi R_{Au}^2 \sim 10\ee
 The maximal number  in the whole fireball  is
 perhaps  an order of magnitude larger \cite{Shuryak:2003sb}. Assuming random statistics,
this implies that the fluctuations of the topological charge in the the fireball is $\Delta Q_{top}\sim \sqrt{N_{sphalerons}}=3-10$. As the size of the sphalerons is $\sim \rho \ll R_{Au}$, each of them
can be viewd as exploding shells. In QGP those evolve diffusively and merge, 
in a process not yet understood quantitatively, producing at the time of fireball freezeout some
chirally asymmetric spots, such as one shown in Fig.(\ref{fig1}).

   Further development was related with the presence of a magnetic field $\vec B$:
   the so called Witten effect states that  when the nonzero $\theta$ is present, it mixes
   with the electric field $\vec E\sim \theta \vec B $. This leads to a current and eventually
 charge separation in the direction of $\vec B$ . The microscopic dynamical mechanism underlying this process has been first explored in \cite{Faccioli}, and has lead to the discussion of charge asymmetry in large bodies \cite{FaccioliMillo} such as neutron stars, Galaxies and  large scale cosmological magnetic field domains. 
 
Kharzeev et al \cite{Kharzeev:2007jp} proposed to look for charge asymmetry
induced by chiral CP-odd fluctuations at
 Relativistic Heavy Ion Collider (RHIC), by means of charged pions' correlations. They pointed out
 that   very intense magnetic field ---up to $10^{19}$Gauss--- is generated by the nuclei and also by
 motion of excited matter, which has certain overall positive charge density and vorticity,
 except in the case of exactly central collisions.
The asymmetry in charged pion correlations relative to the collision plane
 has been recently observed by the Star Collaboration \cite{star1} \cite{star2}, see Fig.(\ref{fig_asymmetry}). The observed effect is indeed absent for central collisions. Still, the 
observed correlation is by itself CP-even $\sim <\theta^2>$ 
  and  in principle it may be induced by some other effect, absent in the usual event generators.  To be sure that it is indeed proportional to $<\theta^2>$ (averaging over the volume and events) some other experimental confirmation would be desirable. Dividing it out would allow to single out 
  the QGP response to $\theta$, a property opening a window to new transport coefficients
  \cite{Son:2009tf,Lublinsky:2009wr}.
  In this paper we propose CP-odd decays as such possibility.

\section{CP forbidden decays}

In this paper we are discussing the CP-violating processes in the phase of heavy ion collisions at the freezeout, which occurs at temperature $T\simeq170$MeV. The system is in the hadronic phase, but the topological fluctuations have not washed out yet and there is a domain of finite size where the QCD vacuum is in a state with an effective $\theta\neq0$. In this regime one can use Chiral Perturbation Theory (ChPT),  keeping in mind that the coefficients of the Chiral Lagrangian  change their values
with temperature . For our purposes we only need the CP-odd part of the ChPT Lagrangian in the leading order in $\theta$.

The ChPT Lagrangian is usually represented as a function of a non-linear field $U(x)~=~e^{\frac{i}{F_\pi}\Pi[x]}$, where $\Pi$ is the nonet of pseudoscalar mesons
\be
\Pi[x] \!\! =
  \!\!\left(
\begin{array}{ccc}
\pi^0+\frac{1}{\sqrt{3}}\varphi^8+\sqrt{\frac{2}{3}}\varphi^0 & \!\!  \!\! \!\! \!\! \!\!\!\!\sqrt{2}\pi^+& \sqrt{2}K^+\\ 
  \!\! \!\! \!\! \!\! \!\!\!\! \!\! \!\! \!\! \!\!\sqrt{2}\pi^- &  \!\! \!\! \!\! \!\! \!\!\!\! \!\! \!\! \!\! \!\! -\pi^0+\frac{1}{\sqrt{3}}\varphi^8+\sqrt{\frac{2}{3}}\varphi^0 &   \sqrt{2}K^0 \\
 \!\! \!\! \!\! \!\! \!\! \!\! \!\! \!\! \!\! \!\!\sqrt{2}K^- &    \!\! \!\!\!\! \!\! \!\! \!\! \!\! \!\!\!\! \!\! \!\! \!\!\!\!\sqrt{2}\bar{K}^0 &   \!\! \!\! \!\! \!\! \!\! \!\! \!\! \!\! \!\!\!\!-\frac{2}{\sqrt{3}}\varphi^8 +\sqrt{\frac{2}{3}}\varphi^0\nonumber
\end{array}
\right)
\ee
The singlet and octet fields
$\phi^0$ and $\phi^8$ are related to the physical $\eta,\eta'$ fields via the mixing angle $\alpha_M$ as
\be
\eta&=&\varphi^8\cos\alpha_M-\varphi^0\sin\alpha_M\\
\eta^{'}&=&\varphi^8\sin\alpha_M+\varphi^0\cos\alpha_M.
\ee
The CP-odd part of the ChPT Lagrangian in the leading chiral order and at leading order in the $1/N_c$ expansion,
 is equal to\cite{pich}
\be
L_{eff}=-i\frac{\chi_{top}N_f}{2N_c}\bar\theta\left[\mbox{Tr}[U(x)-U^\dagger(x)]-2\log\det U(x)\right]
\label{Leff}
\ee
where $\bar\theta$ is an {\it effective $\theta$ parameter}.
\be\bar\theta\simeq\frac{F_\pi^2 N_c m_\pi^2}{4 N_f  \chi_{top}}\theta\ee
To study decay processes of a pseudoscalar meson into two pseudoscalar mesons, we then have to expand Eq.(\ref{Leff}) to third order in the meson fields and select all the interaction terms which are compatible with a decay process, that is to say terms with a field of rest mass bigger than the sum of the rest masses of the two other fields. A simple calculation reveals that the only decays allowed are $\eta$'s into two pions
\be
&&L_{\eta\rightarrow\pi\pi}=-\frac{2}{\sqrt{3}}\frac{\chi_{top}N_f}{F_\pi^3N_c} \bar{\theta} \times\nonumber\\
&\times&\bigg\{\eta\left(\frac{1}{2}\pi^0\pi^0+\pi^+\pi^-\right)\left[\cos\alpha_M-\sqrt{2}\sin\alpha_M\right]\nonumber\\
&+&\eta^{'}\left(\frac{1}{2}\pi^0\pi^0+\pi^+\pi^-\right)\left[\sqrt{2}\cos\alpha_M+\sin\alpha_M\right]\bigg\}
\ee   
and we will only consider the process $\eta/\eta^{'}\rightarrow\pi^+\pi^-$ because it is simpler to detect charged pions.
The amplitude of this process at leading order is given by the square of the coefficient of the corrispondent term in the Lagrangian
\be
|A_{\eta\rightarrow\pi^+\pi^-}|^2  &=&\frac{4}{3}\frac{\chi_{top}^2N_f^2}{F_\pi^6N_c^2}\bar{\theta}^2\left[\cos\alpha_M-\sqrt{2}\sin\alpha_M\right]^2\nonumber\\
&\simeq&2.37\frac{\chi_{top}^2}{F_\pi^6}\bar{\theta}^2\\
|A_{\eta^{'}\rightarrow\pi^+\pi^-}|^2  &=&\frac{4}{3}\frac{\chi_{top}^2N_f^2}{F_\pi^6N_c^2}\bar{\theta}^2\left[\sqrt{2}\cos\alpha_M+\sin\alpha_M\right]^2\nonumber\\
&\simeq&1.63\frac{\chi_{top}^2}{F_\pi^6}\bar{\theta}^2\ee
for value of the mixing angle we used the value $\alpha_M\simeq -15^\circ$, see \cite{mixing}. The topological susceptibility and the pion decay constant have to be substituted with their values at the freezout temperature $T_f\simeq170$MeV.

The decay probability is then equal to
\be
\Gamma_{\eta\rightarrow\pi^+\pi^-}&=&\frac{1}{16\pi}|A_{\eta\rightarrow\pi^+\pi^-}|^2  \frac{\sqrt{m_\eta^2-4m_\pi^2}}{m_\eta^2}\nonumber\\
&\approx&  127 MeV \bar\theta^2\\
\Gamma_{\eta^{'}\rightarrow\pi^+\pi^-}&=&\frac{1}{16\pi}|A_{\eta^{'}\rightarrow\pi^+\pi^-}|^2  \frac{\sqrt{m_{\eta^{'}}^{2}-4m_\pi^2}}{m_{\eta^{'}}^{2}}\nonumber\\
&\approx&80MeV\bar\theta^2
\ee
where in the last approximate equality we have substituted the numerical values of the parameters involved, at $T=0$. (We return below to the issue of $T$ dependence).

We now make a comparison to allowed decays: known decays into $\eta\rightarrow 3\pi$ are so much suppressed that the corresponding partial widths
 are about half of the total width of $\Gamma_\eta=1.18\pm.11\, keV$. So, with $\langle\theta^2\rangle\sim 10^{-2}-10^{-3}$  needed to explain CP-odd fluctuations at RHIC, we actually find that
 the forbidden decay rate in the asymmetric spots is orders of magnitude $larger$ than the allowed one! Still, the corresponding lifetime is much larger than the duration of the freezeout  $\Delta \tau_f$, after which all particles including etas
 leave the spot.  Thus the fraction of etas to decay via the forbidden channel is still small
 \be P_{\eta\rightarrow\pi^+\pi^-}=\Gamma_{\eta\rightarrow\pi^+\pi^-}\Delta \tau_f{V_{spot} \over V_f}\sim  0.1 \theta^2 {\Delta \tau_f \over 1 fm}{V_{spot} \over V_f}
 \ee
of the order of one per mill or so. (The rest decay in the usual way outside the spot.) 

We now discuss how masses and parameters of the chiral Lagrangian (like topological susceptibilties) are modified from their vacuum values, at finite temperature. We start by considering the mass of eta, which is given by Gell-Mann-Oakes-Renner relation and thus
\be m_\eta^2(T)\sim m_s {<\bar s s>(T) \over f_\eta^2(T)}.\ee
While the quark condesate changes appriciably for $T\rightarrow T_f$, the same is true for $f_\pi,f_\eta$, and the ratio is believed to be changed by very small amount. We looked at available lattice literature,
and although there are measurements of all three quantities $m_\eta(T),<\bar s s>(T) , f_\pi(T)$
we have not found common datasets in which all can be compared together. Lacking that,
uncertainties due to different lattices and quark masses are too large to give some
particular numbers.  Yet nothing prevents to do so in a dedicated calculation.

The topological susceptibility does not change appreciabily and account just for a small factor in the
$eta$ mass: it may however change  $\Gamma_{\eta\rightarrow\pi\pi}$ and -- more importantly -- the 
value of the $\eta'$ mass. 

We will now subsequently discuss the factors which affect the invariant mass window in which decays are to be observed. These include: \\
(i) spatial dependence of effective $\theta$ parameter\\
(ii) meson mass modification as a function of temperature.

(i) The spatial dependence of the effective parameter $\theta(x)$
can be viewed as additional 3-momentum $\vec k\approx 1/R_{spot}$ coming from its Fourier
transform. (We do not include the time-dependent component, assuming
that diffusion of chirality is a slow process, and the correspondig
frequency is negligible.) As a result, the invariant mass of the 
final state is modified by
\be m_{\eta} = \sqrt{m_\eta^2-\vec k^2}\approx m_\eta\bigg[1-{\vec k^2\over 2 m_\eta^2}\bigg] \ee 
The magnitude of the spot size and the corresponding momenta will be discussed in the next section, for estimate we will now use $k=50-100 \, MeV$.
From the formula above we see that the correction to the mass is then 
in the range of $-(0.005\div0.02)$. 

(ii) thermal mass modification of the eta, can also be viewed as
the energy shift of a particles sitting in a certain potential created by all other memebers of the fireball, at the moment of the decay. In our case we cosider two different inital particles, $\eta,\eta'$, which are related with
 different phenomena and thus should experience different shifts. The former one is mostly Goldstone meson (ignoring their mixing), the SU(3) partner
of the pions, thus its mass is obtained from Gell-Mann-Oaks-Renner
relation; the latter, $\eta^{'}$, has a mass given by Witten-Veneziano formula.
While the quark condensate and the decay constant are expected to have
significant changes themselves, and in the chiral limit vanish at $T=T_c$,
the ratio appearing in the eta mass is believed to change with $T$ much less.
For example, the quark condensate at $T_{ch}\approx 170\, MeV$
is reduced by about 20\% \cite{Bazavov:2009zn}.

At the end of this section let us briefly discuss 
 the relaxation of the spot itself. During the freezeout period, when the CP-odd spot is populated with mesons,
 some energy/momentum exchange between $\theta(t,x)$ and these mesons is possible and is described by the same chiral Lagrangian:
 for example, it was included in our discussion of the visible width of the forbidden decays above. What happens after that period, when
 the residual  $\theta(t,x)$ remains alone, as a vacuum perturbation? Depending on the energy available, it can still decay into a number of channels.
 In the order of reduced mass those would be e.g. processes like 
   $\theta\rightarrow \eta'$, or  $\theta\rightarrow \eta\pi\pi$ or eventually  $\theta\rightarrow \gamma\gamma$ till it relaxes completely to $\theta=0$.
   The rate of all of them, if needed, can be calculated from the same standard chiral Lagrangian.

\section{What is the size of the ``chirally asymmetric spot"?}
\label{sec_size}
\begin{figure}[!t]
\vspace{0.75cm}
\includegraphics[width=8cm]{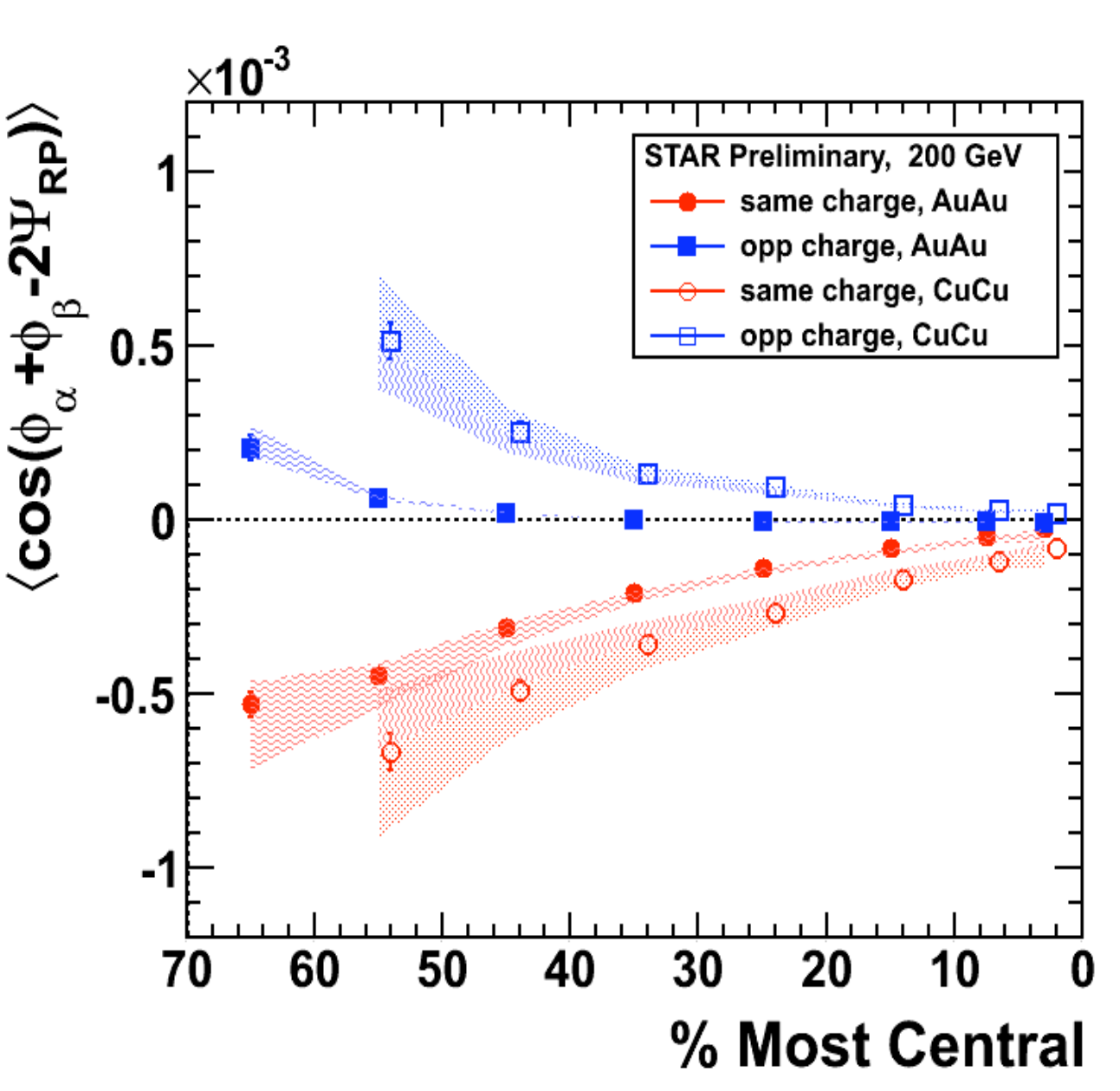}
\caption[h]{\label{fig_asymmetry} 
Preliminary STAR data on charge correlation \protect\cite{star1,star2}.}
\end{figure}

Let us return to the geometry of the charge correlations
 shown in fig.\ref{fig_asymmetry}. Note that we placed the asymmetric spot  at the periphery of the fireball:
we did so for the following reason (known as trigger bias): such geometry increases
chances that $both$ charge pions (say $2\pi^+$) 
 are observed, escaping from the system unscattered and moving in a certain direction (up
 in the figure). There is no such
  bias  effect for opposite charges $+-$  since those are expected to move approximately
  in opposite direction. Existing experience with jets show that one of the two or both
  would be rescattering and thus quenched by matter much more.
  So, the observed
ratio of opposite-to-same charge effects $R(+-/++)$ is expected to be small.

 Now look at the STAR data shown in fig.\ref{fig_asymmetry}. Those for AuAu collisions   (shown by the closed points) indeed finds  the effect for opposite charge $+-$ correlation to be much smaller than for same charge $++/--$ one, $R(+-/++)\ll 1$. However more recent data taken for CuCu collisions
 (open points) show   a nearly  symmetric picture, with $|R(+-/++)|\approx  1$ (up to a sign, as expected).
 We take it as the direct experimental evidence for the fireball sizes at freezeout having the following relation to the size of the CP-odd spot 
 \be R_{AuAu} \gg R_{spot}\sim R_{CuCu} \ee  
We used this last approximate equality for the estimate of the corresponding momentum in the previous secton.

\section{Summary and Discussion}
 The 
 charge pion correlations observed by STAR collaboraton \cite{star1} \cite{star2}
 is by itself CP-even. 
   To be sure that it is indeed proportional to $<\theta^2>$, we propose to look for 
   CP-forbidden decays such as $\eta\rightarrow \pi^+\pi^-, \eta'\rightarrow \pi^+\pi^-$.
   According to our estimates, the rates of the decays at freezeout are orders of magnitude
   larger than the allowed 3-meson modes, and the total fraction of the forbidden decays
   should be of the order of $10^{-3}$ or so. 
   
   We think that with the available statistics of $\sim 10^9$ events one can separate the
   peaks in the invariant mass from the statistical background of random $\pi^+\pi^-$ 
pairs with the same invariant mass.
   (Since the channels are very different, the usual decays
make no background.) 
From theoretical point of view $\eta$ stands a better chance because its mass is expected to
change little, as it is proportional to the square root of the ratio of quark condensate and $F_\pi$,
with similar $T$-dependence. The $\eta'$ mass is related to topological susceptibility and
may change more. But from experimental point of view,
   due to the values of $\eta$ and $\eta^{'}$ rest mass, the signal/background ratio may be larger for $\eta^{'}$ decays: since the freezeout temperature it is $T_f\simeq170MeV$, RHIC collisions are more likely to produce pions with momentum of the order of $0.5\,m_\eta$ than $0.5\,m_\eta^{'}$.
   The issue of course require further studies, including experimental resolution and acceptance:
   but, we repeat, it does not require new data, just their analysis.
   
   If observed, the CP-forbidden decays would provide 
   an independent measure of the CP-odd fluctuations. Dividing $<\theta^2>$
    out would allow to single out 
  the QGP response to $\theta$, a property opening a window to new transport coefficients
  \cite{Son:2009tf,Lublinsky:2009wr}.

\section*{Acknowledgments}
The work of RF is supported in part by the physics department of the University of Trento and INFN. The work of ES is supported in parts by the US-DOE grant DE-FG-88ER40388.


\begin{thebibliography}{99}
\bibitem{Belavin:1975fg}
  A.~A.~Belavin, A.~M.~Polyakov, A.~S.~Shvarts and Yu.~S.~Tyupkin,
  Phys.\ Lett.\  B {\bf 59}, 85 (1975).

\bibitem{Shuryak:1981ff}
 E.~V.~Shuryak,
 Nucl.\ Phys.\  B {\bf 203}, 93 (1982).

\bibitem{Schafer:1996wv}
 T.~Schafer and E.~V.~Shuryak,
 Rev.\ Mod.\ Phys.\  {\bf 70}, 323 (1998)
 [arXiv:hep-ph/9610451].

\bibitem{Kharzeev:2000ef}
 D.~E.~Kharzeev, Y.~V.~Kovchegov and E.~Levin,
 Nucl.\ Phys.\  A {\bf 690}, 621 (2001)
 [arXiv:hep-ph/0007182].

\bibitem{Shuryak:2000df}
 E.~V.~Shuryak and I.~Zahed,
 Phys.\ Rev.\  D {\bf 62}, 085014 (2000)
 [arXiv:hep-ph/0005152].

\bibitem{Ostrovsky:2002cg}
 D.~M.~Ostrovsky, G.~W.~Carter and E.~V.~Shuryak,
 Phys.\ Rev.\  D {\bf 66}, 036004 (2002)
 [arXiv:hep-ph/0204224].

\bibitem{Shuryak:2002qz}
 E.~Shuryak and I.~Zahed,
 Phys.\ Rev.\  D {\bf 67}, 014006 (2003)
 [arXiv:hep-ph/0206022].

\bibitem{Shuryak:2003sb}
 E.~V.~Shuryak,
 Nucl.\ Phys.\  A {\bf 717}, 291 (2003).

\bibitem{Kharzeev:1998kz}
 D.~Kharzeev, R.~D.~Pisarski and M.~H.~G.~Tytgat,
 Phys.\ Rev.\ Lett.\  {\bf 81}, 512 (1998)
 [arXiv:hep-ph/9804221].

\bibitem{Lucini:2004yh}
 B.~Lucini, M.~Teper and U.~Wenger,
 Nucl.\ Phys.\  B {\bf 715}, 461 (2005)
 [arXiv:hep-lat/0401028].

\bibitem{Faccioli}
 P.~Faccioli
 Phys.\ Rev.\ D {\bf 71},  091502 (2005)

\bibitem{FaccioliMillo}
 R.~Millo, P.~Faccioli
 Phys.\ Rev.\  D {\bf 77}, 065013 (2008)
 [arXiv:hep-ph/0706.0805]. 

\bibitem{star1}
 B.~I.~Abelev et al. [The STAR Collaboration]
 Phys.\ Rev.\ Lett., {\it to appear}
 [arXiv:nucl-ex/0909.1739]

\bibitem{star2}
 B.~I.~Abelev et al. [The STAR Collaboration]
 [arXiv:nucl-ex/0909.1717]

\bibitem{Kharzeev:2007jp}
 D.~E.~Kharzeev, L.~D.~McLerran and H.~J.~Warringa,
 Nucl.\ Phys.\  A {\bf 803}, 227 (2008)
 [arXiv:0711.0950 [hep-ph]].

\bibitem{Son:2009tf}
  D.~T.~Son and P.~Surowka,
  Phys.\ Rev.\ Lett.\  {\bf 103}, 191601 (2009)
  [arXiv:0906.5044 [hep-th]].
\bibitem{Lublinsky:2009wr}
  M.~Lublinsky and I.~Zahed,
  arXiv:0910.1373 [hep-th].

 \bibitem{pich}
 A.~Pich, E.~De Rafael
 Nucl.\ Phys.\ B{\bf 367}, 313-333, (1991)


 \bibitem{mixing}
 A.~Bramon, R.~Escribano, M.~D.~Scadron
 Eur.\ Phys.\ Journ. C\bf{7}: 271-278 (1999)




\bibitem{Bazavov:2009zn}
 A.~Bazavov {\it et al.},
 Phys.\ Rev.\  D {\bf 80}, 014504 (2009)
 [arXiv:0903.4379 [hep-lat]].



\end{thebibliography}
\end{document}